\documentclass{PoS}

\usepackage[utf8]{inputenc}
\usepackage{amsmath}
\usepackage{amssymb}
\usepackage{color}
\usepackage{graphicx}
\usepackage{multirow}

\usepackage{subcaption}
\usepackage{braket}

\title{Exploring possibly existing $q q \bar b \bar b$ tetraquark states with $q q
= ud, ss, cc$}

\ShortTitle{Exploring possibly existing $q q \bar b \bar b$ tetraquark states with $q q
= ud, ss, cc$}

\author{\speaker{Antje Peters}$^a$, Pedro Bicudo$^b$, Krzysztof Cichy$^{a,c}$,  Bj\"orn Wagenbach$^a$, Marc Wagner$^a$ \\

$^a$Goethe-Universit\"at Frankfurt am Main, Institut f\"ur Theoretische Physik,\\
Max-von-Laue-Stra{\ss}e 1, D-60438 Frankfurt am Main, Germany\\

$^b$CFTP, Departamento de Física, Instituto Superior Técnico, Universidade de Lisboa,\\
Avenida Rovisco Pais, 1049-001 Lisboa, Portugal\\

$^c$Adam Mickiewicz University, Faculty of Physics,\\
 Umultowska 85, 61-614 Poznan, Poland\\

E-mail: \email{peters@th.physik.uni-frankfurt.de, bicudo@tecnico.ulisboa.pt, kcichy@th.physik.uni-frankfurt.de,  wagenbach@th.physik.uni-frankfurt.de, mwagner@th.physik.uni-frankfurt.de}
}

\abstract{
We compute potentials of two static antiquarks in the presence of two quarks $qq$ of finite mass using lattice QCD. In a second step we solve the Schr\"odinger equation, to determine, whether the resulting potentials are sufficiently attractive to host a bound state, which would indicate the existence of a stable $q q \bar b \bar b$ tetraquark. 
We find a bound state for $qq=(ud-du)/\sqrt{2}$ with corresponding quantum numbers $I(J^ P)=0(1^+)$ and evidence against the existence of bound states with isospin $I=1$ or $qq \in \{cc,ss \}$.
}

\FullConference{The 33rd International Symposium on Lattice Field Theory\\
                 14 -18 July  2015\\
                 Kobe International Conference Center, Kobe, Japan}

\begin{document}


\section{Motivation}

A number of mesons observed in experiments like LHCb or Belle are not well understood. Those mesons have masses and quantum numbers, which are not typical for standard quark-antiquark states, but indicate an exotic four-quark structure. Prominent examples are the charged charmonium-like and bottomonium-like states ${Z_c}^\pm$ and ${Z_b}^\pm$ (cf.\ e.g.\ \cite{Belle:2011aa}). Their masses and decay products suggest the presence of a $c \bar c$ or $b \bar b$ pair, respectively. On the other hand their electric charge indicates additionally a light quark-antiquark pair $u \bar d$ or $d \bar u$. Those four-quark systems, in the following also referred to as tetraquarks, are expected to be studied in more detail in the near future by experimental collaborations. Therefore, a sound theoretical understanding of those systems is crucial and of great interest.

Here we summarize the main results of our recently published work \cite{Bicudo:2015vta}, where we have studied four-quark systems with two heavy antiquarks $\bar b \bar b$ and two lighter quarks $q q$ using lattice QCD and the Born-Oppenheimer approximation. First $\bar b \bar b$ potentials in the presence of lighter quarks $q q$ are computed. Then the Schr\"odinger equation is solved using these potentials, where possibly existing bound states indicate stable tetraquarks. Other papers studying the same systems with similar methods include \cite{Stewart:1998hk,Michael:1999nq,Cook:2002am,Doi:2006kx,Detmold:2007wk,Wagner:2010ad,
Bali:2010xa,Wagner:2011ev,Bicudo:2012qt,Brown:2012tm,Wagenbach:2014oxa,
Scheunert:2015pqa}.


\section{Qualitative discussion of $q q \bar b \bar b$ systems}

At small $\bar b \bar b$ separations the $\bar b \bar b$ interaction is dominated by 1-gluon exchange. For a bound state the $\bar b \bar b$ pair must, therefore, be in an attractive color triplet. Due to the Pauli principle and because we assume a spatially symmetric $s$-wave, $\bar b \bar b$ has to form an antisymmetric color-spin-flavor combination and, hence, a symmetric spin combination, i.e.\ $\bar b \bar b$ spin $j_b = 1$. Since the complete four-quark system is color neutral, the light quarks $q q$ must be in an antisymmetric color antitriplet. Again due to the Pauli principle $q q$ has to form an antisymmetric color-spin-flavor combination and, hence, a symmetric spin-flavor combination. Candidates for tetraquarks are, therefore, the (spin) \textit{scalar isosinglet} (i.e.\ a $q q$ spin singlet $j = 0$ with antisymmetric flavor, e.g.\ $qq \in \{ (ud - du) / \sqrt{2} \, , \, (s^{(1)} s^{(2)} - s^{(2)} s^{(1)}) / \sqrt{2} \, , \, (c^{(1)} c^{(2)} - c^{(2)} c^{(1)}) / \sqrt{2} \}$\footnote{To be able to study flavor antisymmetric $q q$ combinations with $q = s$, we consider two hypothetical degenerate flavors with the mass of the $s$ quark, $s^{(1)}$ and $s^{(2)}$, and similarly for $q = c$, $c^{(1)}$ and $c^{(2)}$.}) and  the (spin) \textit{vector isotriplet} (i.e.\ a $q q$ spin triplet $j = 1$ with symmetric flavor, e.g.\ $qq \in \{ uu \, , \, (ud + du) / \sqrt{2} \, , \, dd \, , \, ss \, , \, cc \}$). The overall quantum numbers of a bound $q q \bar b \bar b$ system are $I(J^P) = 0(1^+)$ for the scalar isosinglet channel and $I(J^P) \in \{ 1(0^+) \, , \, 1(1^+) \, , \, 1(2^+) \}$ for the vector isotriplet channel.

At large $\bar b \bar b$ separations the $\bar b \bar b$ interaction is screened by the light quarks $q q$, i.e.\ the four quarks form a system of two heavy-light mesons. One expects stronger screening for increasing quark mass $m_q$, because the wave functions of the corresponding mesons $q \bar b$ are then more compact.


\section{Lattice QCD computation of static antiquark-antiquark potentials}

We extract potentials of two static antiquarks $\bar Q \bar Q$ (approximating the two $\bar{b}$ quarks of the $q q \bar b \bar b$ system) in the presence of two light quarks $q q$ from correlation functions
\begin{equation}
\label{corr} C(t,r) \ \ = \ \ \bra{\Omega} \mathcal O^\dagger(t) \mathcal O(0) \ket{\Omega} \ \underset{t \to \infty}{\propto} \ \exp(-V(r) t) .
\end{equation}
$\mathcal O$ denotes a four-quark creation operator,
\begin{equation}
\label{EQN001} \mathcal O \ \ = \ \ (\mathcal{C} \Gamma)_{AB} (\mathcal{C} \tilde{\Gamma})_{CD} \Big(\bar{Q}_C(\mathbf{r}_1) q_A^{(1)}(\mathbf{r}_1)\Big) \Big(\bar{Q}_D(\mathbf{r}_2) q_B^{(2)}(\mathbf{r}_2)\Big) \quad , \quad r \ \ = \ \ |\mathbf{r}_1 - \mathbf{r}_2| ,
\end{equation}
where $\Gamma$ is an appropriate combination of $\gamma$ matrices accounting for defined quantum numbers light quark spin $|j_z|$, parity $P$ and $P_x$ (cf.\ \cite{Wagner:2010ad} for details). $\Tilde \Gamma \in \{ (1-\gamma_0)\gamma_5 \, , \, (1-\gamma_0) \gamma_j \}$ does not affect the resulting potential $V(r)$, since the static quark spin is irrelevant. $\mathcal{C} = \gamma_0 \gamma_2$ denotes the charge conjugation matrix. Note that operators like (\ref{EQN001}) generate overlap not only to mesonic molecule structures, but also to diquark-antidiquark structures \cite{Alexandrou:2012rm,Abdel-Rehim:2014zwa}.

The asymptotic value of a potential and whether it is attractive or repulsive depends on the quantum numbers $(|j_z|,P,P_x)$ and, hence, on $\Gamma$. In the following we are exclusively interested in attractive potentials between two ground state static-light mesons: the scalar isosinglet corresponding to $\Gamma = (1 + \gamma_0) \gamma_5$ and the vector isotriplet corresponding to $\Gamma = (1 + \gamma_0) \gamma_j$.

Computations have been performed using two ensembles of gauge link configurations generated by the European Twisted Mass Collaboration (ETMC) with dynamical $u/d$ quarks. Information on these ensembles can be found in Table~\ref{tab:setup} and \cite{Boucaud:2008xu,Baron:2009wt}.

\begin{table}[htb]
\centering
\begin{tabular}{cccccc}
$\beta$ & lattice size & $\mu_l$ & $a$ in fm & $m_\pi$ in MeV & \# configurations \\
\hline
$3.90$ & $24^3 \times 48$ & $0.00400$ & $0.079$ & $340$ & $480$ \\
$4.35$ & $32^3 \times 64$ & $0.00175$ & $0.042$ & $352$ & $100$ \\
\end{tabular}
\caption{\label{tab:setup}Ensembles of gauge link configurations ($\beta$: inverse gauge coupling; $\mu_l$: bare $u/d$ quark mass in lattice units; $a$: lattice spacing; $m_\pi$: pion mass).}
\end{table}


\section{$q q \bar b \bar b$ tetraquarks in the Born-Oppenheimer approximation}

To determine an analytical expression for the $\bar Q \bar Q$ potential or equivalently $\bar b \bar b$ potential, we fit the ansatz
\begin{equation}
\label{fitfunction} V(r) \ \ = \ \ -\frac{\alpha}{r} \exp\bigg(-\bigg(\frac{r}{d}\bigg)^2\bigg) + V_0
\end{equation}
with respect to $\alpha$, $d$ and $V_0$ to the lattice QCD results obtained in the previous section. The constant $V_0$ accounts for twice the mass of the ground state static-light meson.

We insert the the analytical expression (\ref{fitfunction}) in the Schr\"odinger equation for the radial coordinate of the two $\bar b$ quarks (which we assume to be in an $s$-wave),
\begin{equation}
\label{schroedinger} \bigg(-\frac{1}{2 \mu} \frac{d^2}{dr^2} + U(r)\bigg) R(r) \ \ = \ \ E_B R(r)
\end{equation}
with $U(r) = V(r)|_{V_0 = 0}$ and $\mu = m_b/2$ and determine the lowest eigenvalue $E_B$. If $E_B < 0$, the four quarks $q q \bar b \bar b$ can form a tetraquark. If $E_B > 0$, there is no binding, i.e.\ the four-quark system will always be a system of two unbound $B$ mesons. Notice that this so-called Born-Oppenheimer approximation is valid for $m_q \ll m_b$, which is certainly the case for $q \in \{ u \, , \, d \, , \, s \}$ and at least crudely fulfilled for $q = c$.

\begin{figure}[htb]
\includegraphics[width=0.49\linewidth]{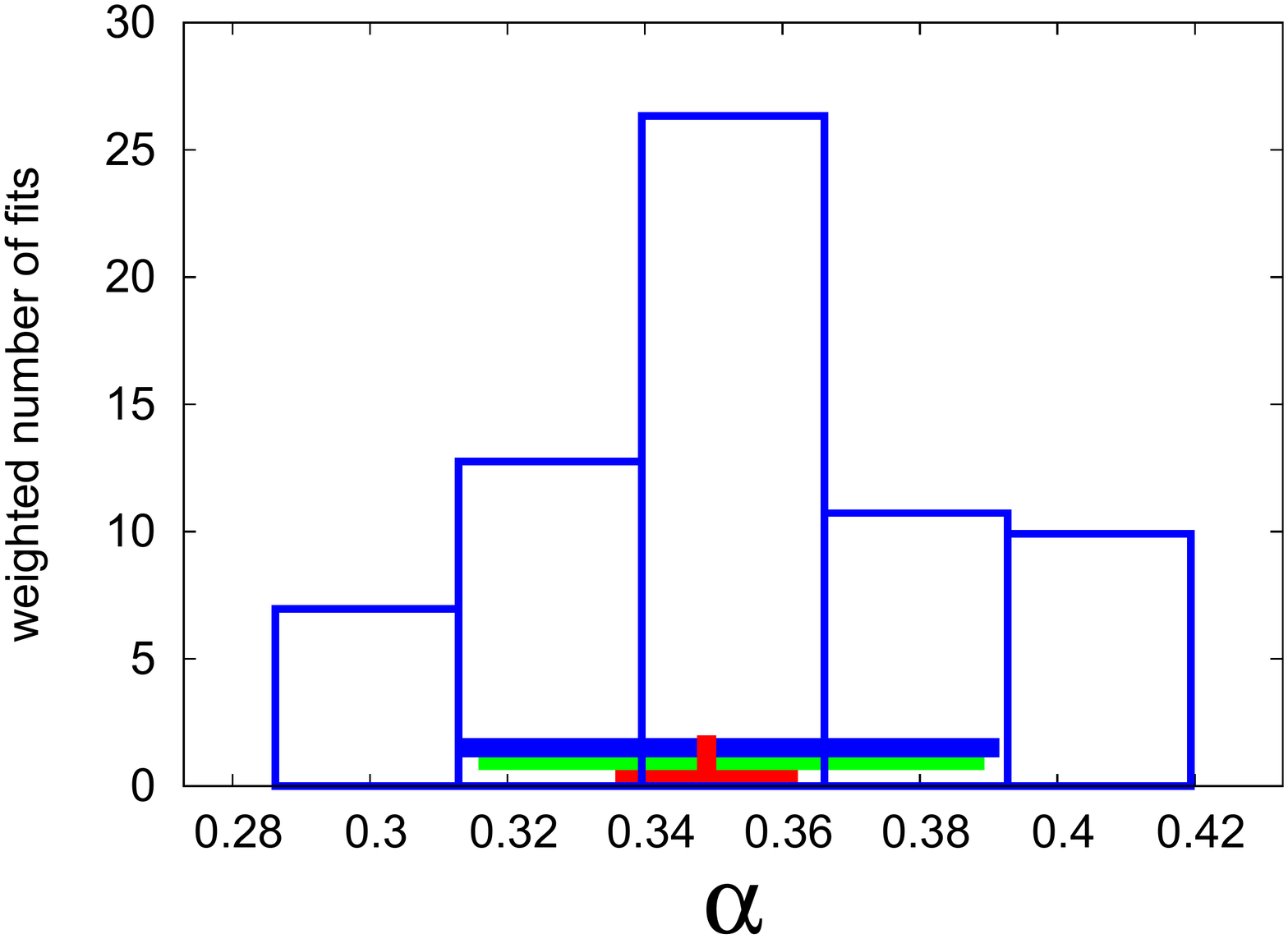}
\includegraphics[width=0.49\linewidth]{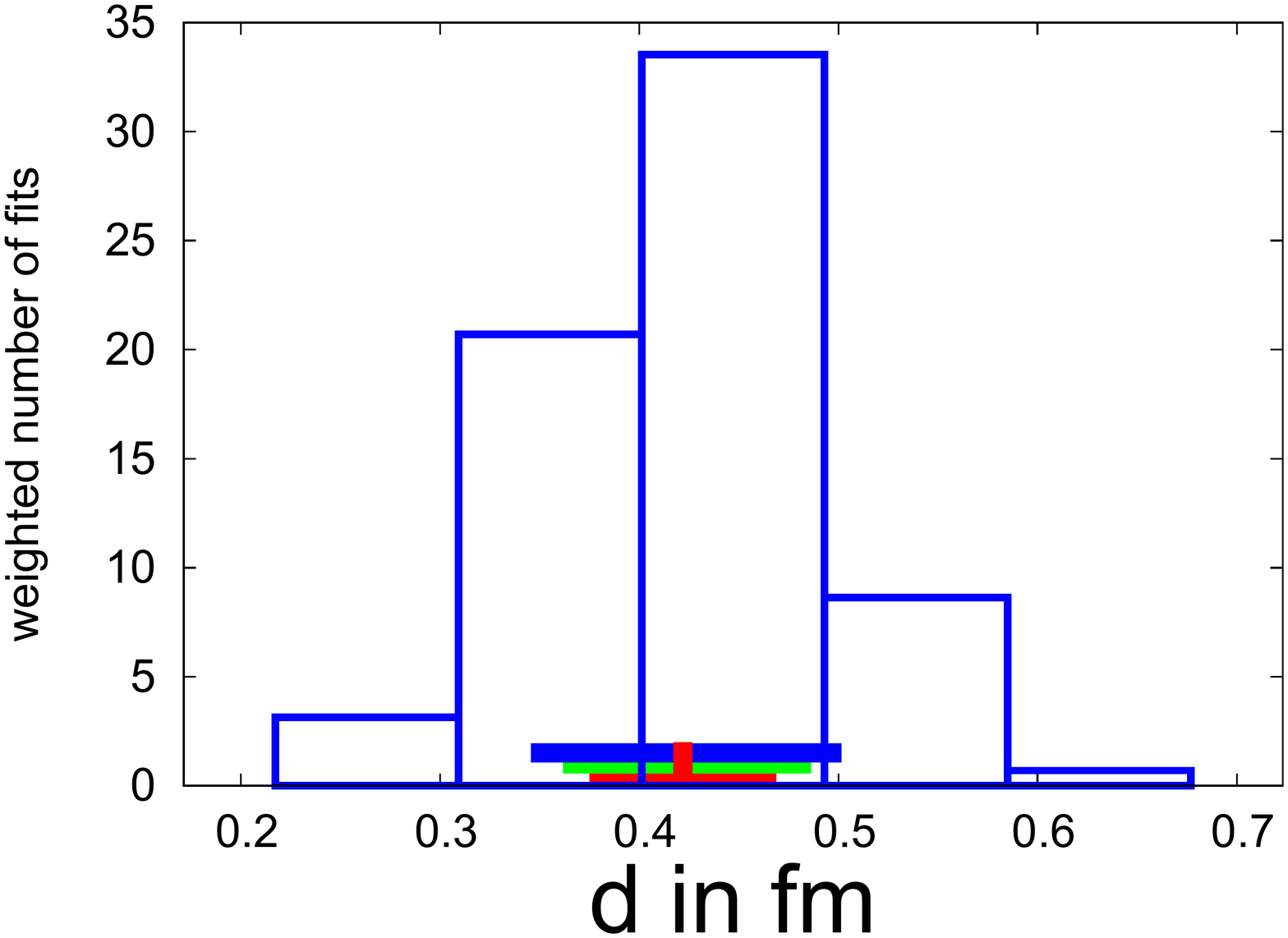}
\caption{\label{histograms}Histograms for the scalar isosinglet for $qq=ud$. The red/green/blue bars indicate the statistical/systematic/combined errors.}
\end{figure}

To quantify the systematic errors of different channels (scalar isosinglet and vector isotriplet, different light flavors $q \in \{ u \, , \, d \, , \, s \, , \, c \}$), we perform a large number of fits varying the range of temporal separations $t_\textrm{min} \leq t \leq t_\textrm{max}$ of the correlation function $C(t,r)$ (cf.\ eq.\ (\ref{corr})), at which the lattice potential is read off, as well as the range of spatial $\bar{b} \bar{b}$ separations $r_\textrm{min} \leq r \leq r_\textrm{max}$ considered in the $\chi^2$ minimizing fit of eq.\ (\ref{fitfunction}) to the lattice potential. Details on this parameter variation can be found in \cite{Bicudo:2015vta}. For each set of input parameters $(t_\textrm{min} , t_\textrm{max} , r_\textrm{min} , r_\textrm{max})$ we determine $\alpha$, $d$ and $E_B$. Then we generate histograms for $\alpha$, $d$ and $E_B$ weighted according to the corresponding $\chi^ 2/\textrm{dof}$. The widths of these histograms are taken as systematic errors of $\alpha$, $d$ and $E_B$ \cite{Cichy:2012vg}, while the statistical errors are obtained via a jackknife analysis. In Figure~\ref{histograms} example histograms for the scalar isosinglet for $qq=ud$ are shown.

\begin{figure}
\includegraphics[width=0.33\linewidth]{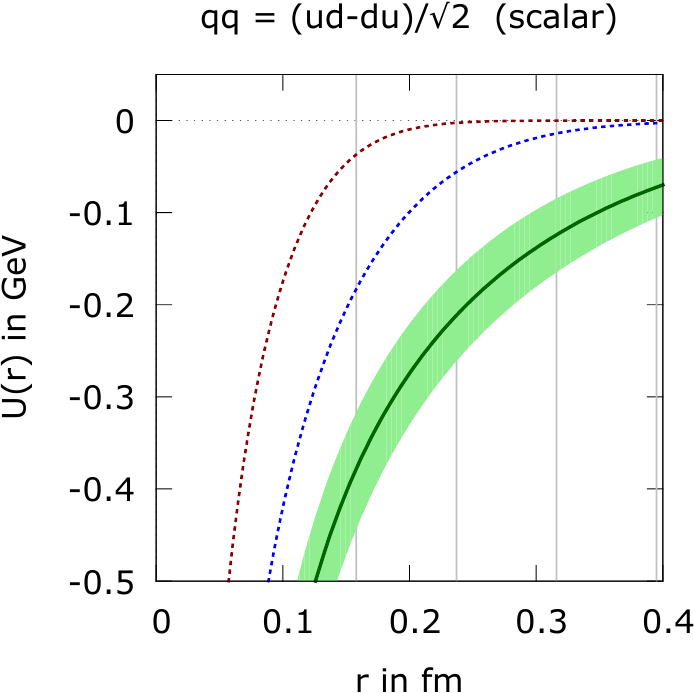}
\includegraphics[width=0.33\linewidth]{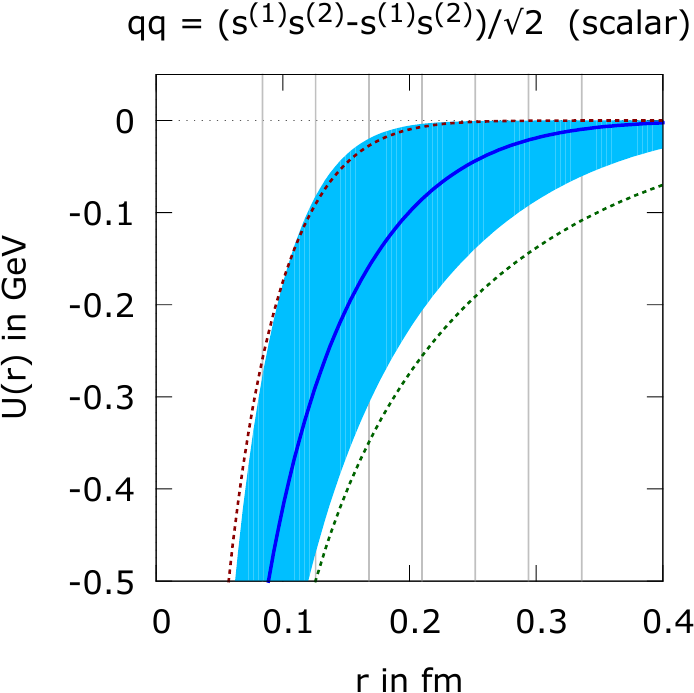}
\includegraphics[width=0.33\linewidth]{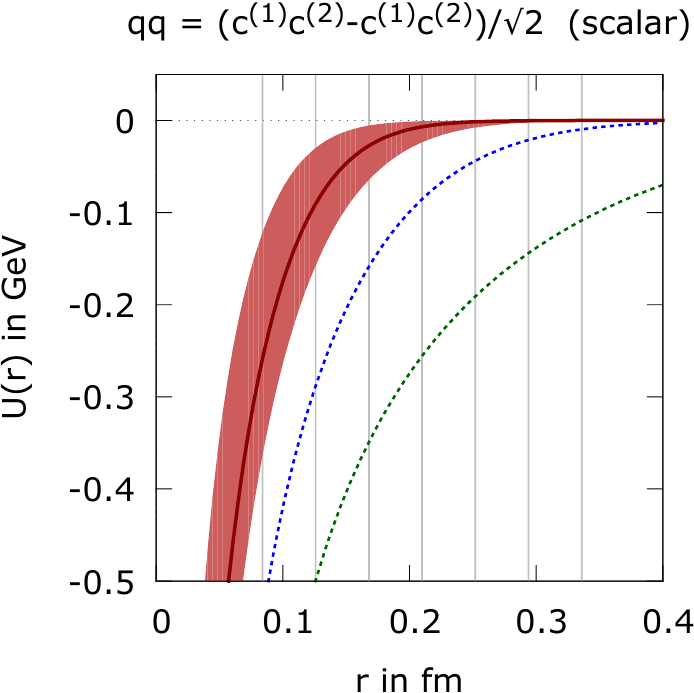} \\
\vspace{-0.2cm} \\
\includegraphics[width=0.33\linewidth]{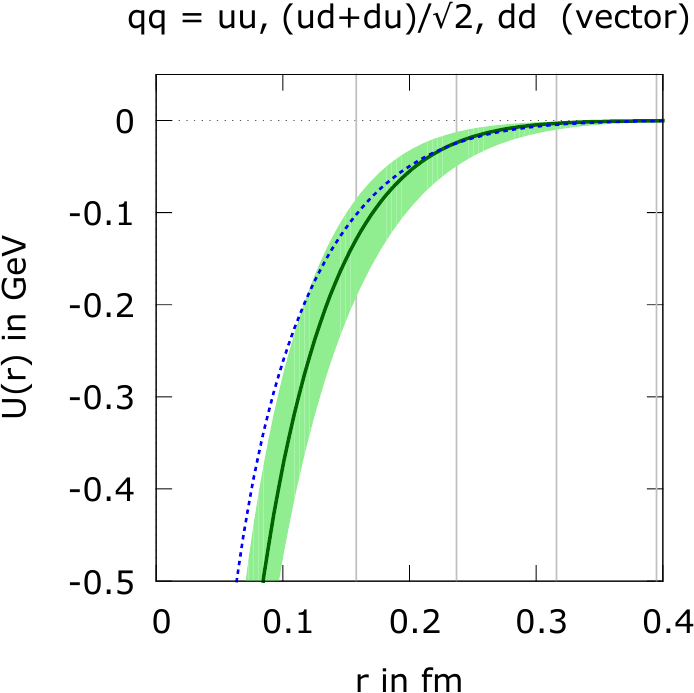}
\includegraphics[width=0.33\linewidth]{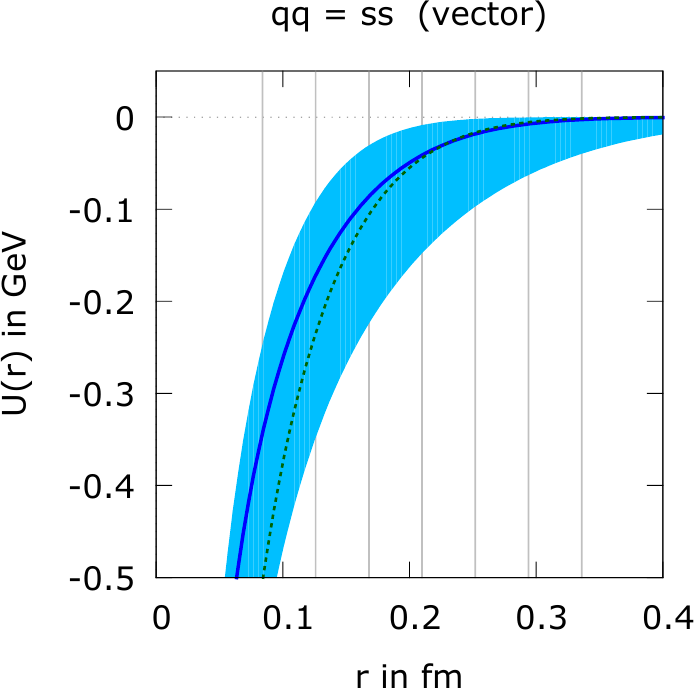}
\phantom{\includegraphics[width=0.33\linewidth]{BILDER/potential_scalar_charm_crp.pdf}}
\caption{\label{allpots}Potentials fits for different channels (upper line: scalar isosinglet; lower line: vector isotriplet). The curves without an error band are copied from the respective other plots in the same line for easy comparison. Vertical lines indicate the available lattice $\bar b \bar b$ separations.}
\end{figure}

The resulting potentials fits for different channels, i.e.\ eq.\ (\ref{fitfunction}) with corresponding values for $\alpha$ and $d$, are collected in Figure~\ref{allpots}. The error bands represent the combined systematic and statistical errors. One can observe that the potentials are wider and deeper for lighter $q q$ quark masses. Moreover, the scalar channels are more attractive than the respective vector channels. Correspondingly, it turns out that there is a bound state only for the scalar isosinglet with $q q = u d$ with binding energy $-E_B = 93_{-43}^{+47} \, \textrm{MeV}$, i.e.\ a bound state with around $2 \sigma$ confidence level.

\begin{figure}[p]
\begin{center}
\includegraphics[width=0.70\textwidth]{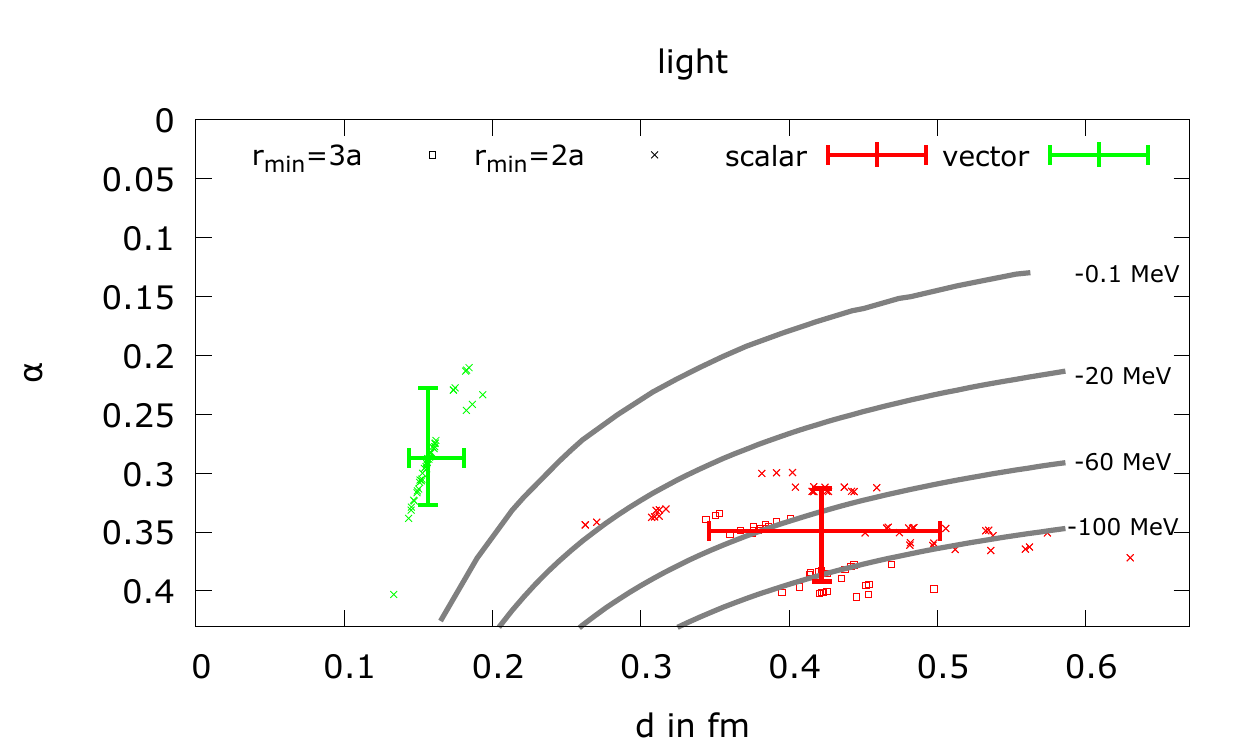} 
\includegraphics[width=0.70\textwidth]{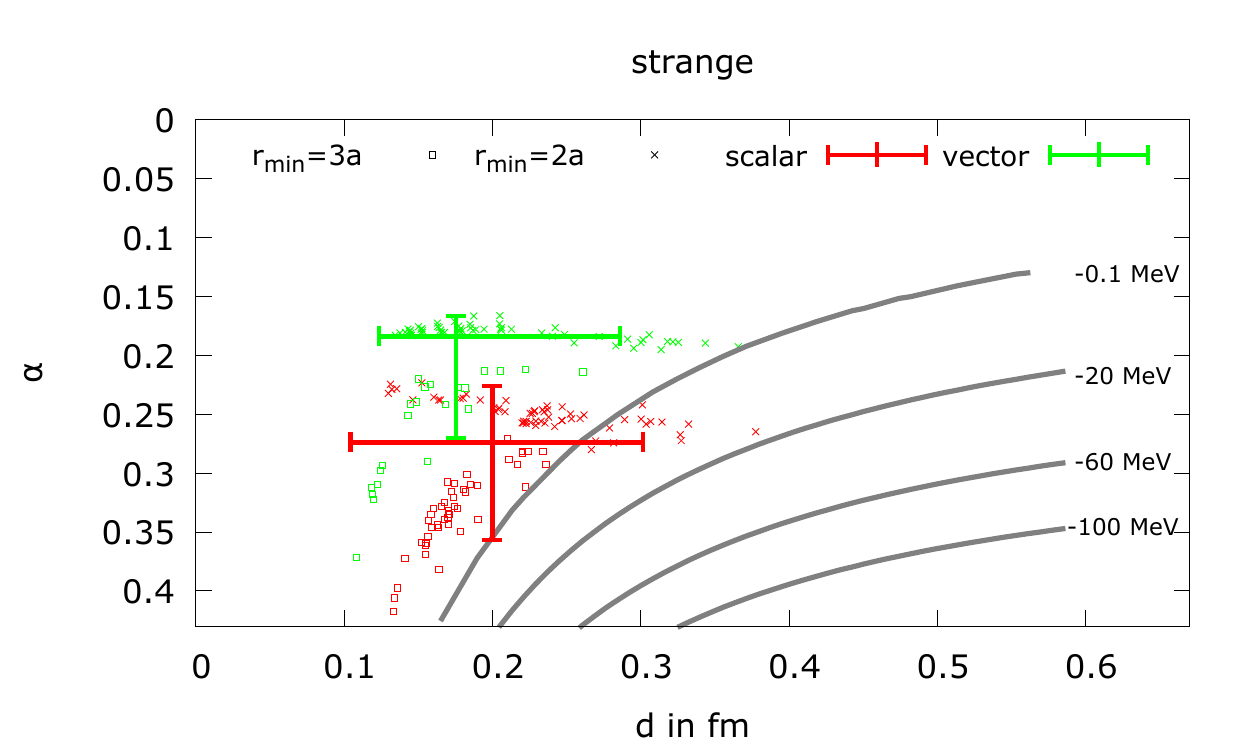} 
\includegraphics[width=0.70\textwidth]{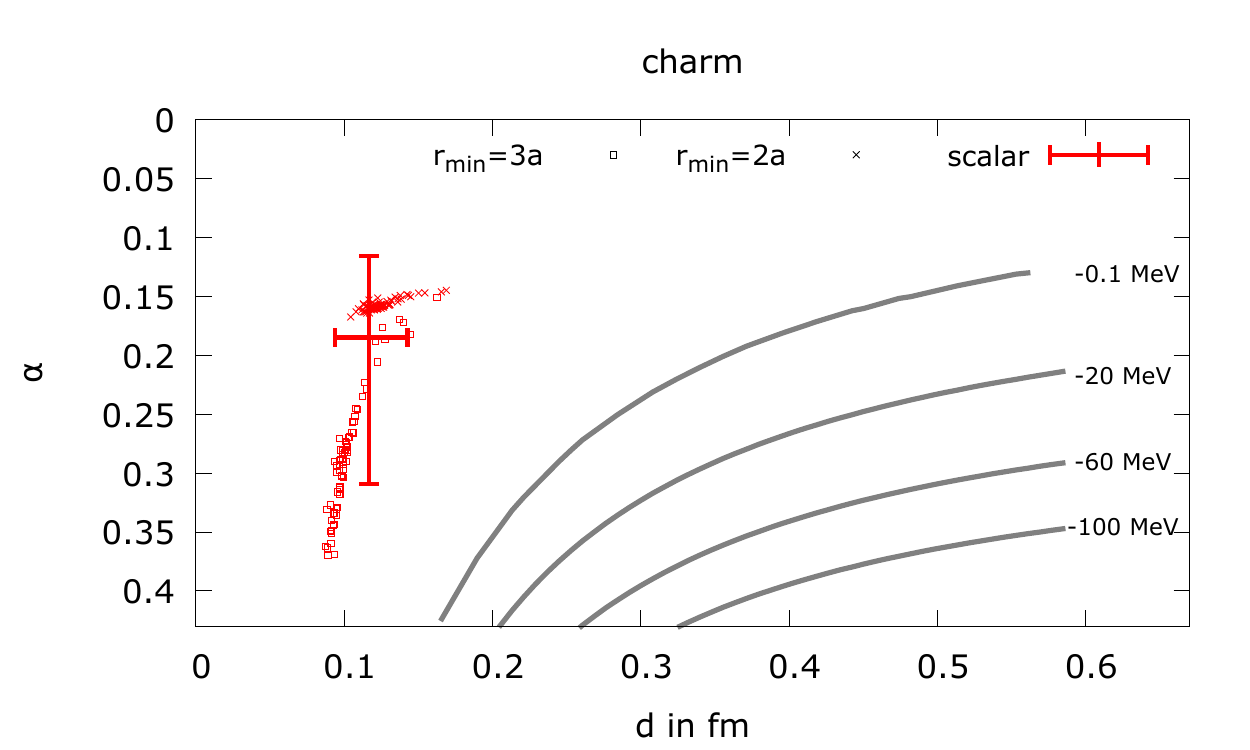}
\end{center}
\caption{\label{FIG340}Binding energy isolines $E_B = \textrm{constant}$ in the $\alpha$-$d$-plane for $u/d$, $s$ and $c$ light quarks $q q$ together with the fit results of eq.\ (4.1) to lattice potentials.}
\end{figure}

In Figure~\ref{FIG340} we present our results in an alternative graphical way. The three plots correspond to $u/d$, $s$ and $c$ light quarks $q q$, respectively. Each fit of eq.\ (\ref{fitfunction}) to lattice potential results is represented by a dot (red: scalar channels; green: vector channels; crosses: $r_\textrm{min} = 2a$;  boxes: $r_\textrm{min} = 3a$). The extensions of the point clouds represent the systematic uncertainties with respect to $\alpha$ and $d$. If a point cloud is localized above or left of the isoline with $E_B = -0.1 \, \textrm{MeV}$ (essentially the binding threshold), the corresponding four quarks $q q \bar{b} \bar{b}$ cannot form a bound state. A localization below or right of that isoline is a strong indication for the existence of a tetraquark. Again there is clear evidence for a tetraquark state in the scalar $u/d$ channel. The scalar $s$ channel is slightly above, but rather close to the binding threshold. The scalar $c$ and all vector channels clearly do not host bound four-quark states.

\section{Summary and outlook}

We have found a $u d \bar b \bar b$ tetraquark with quantum numbers $I(J^P)=0(1^+)$ (i.e.\ in the scalar isosinglet channel with $q q = u d$) with a confidence level of around $2 \sigma$. There seem to exist no tetraquarks for the other channels.

In this work lattice QCD computations have been performed for light $u/d$ quarks corresponding to $m_\pi \approx 340 \, \textrm{MeV}$. We plan to repeat the analysis for at least another pion mass and then extrapolate to the physical point. It will then be most interesting to check, whether a bound state will also appear in the vector isotriplet channel with $q q = ud$. Another aspect is to investigate the structure of the found $I(J^P)=0(1^+)$ tetraquark, i.e.\ to explore, whether it is rather a mesonic molecule or a diquark-antidiquark pair. We also plan to include corrections due to the heavy quark spins (for first preliminary results cf.\ \cite{Scheunert:2015pqa}). Finally, one should study the experimentally more accessible, but theoretically more challenging case of $q \bar q b \bar b$ systems.


\acknowledgments

P.B.\ thanks IFT for hospitality and CFTP, grant FCT UID/FIS/00777/2013, for support. M.W.\ and A.P.\ acknowledge support by the Emmy Noether Programme of the DFG (German Research Foundation), grant WA 3000/1-1.

This work was supported in part by the Helmholtz International Center for FAIR within the framework of the LOEWE program launched by the State of Hesse.

Calculations on the LOEWE-CSC high-performance computer of Johann Wolfgang Goethe-University Frankfurt am Main were conducted for this research. We would like to thank HPC-Hessen, funded by the State Ministry of Higher Education, Research and the Arts, for programming advice.



\end{document}